\documentclass[twocolumn]{aastex631}
\usepackage{amsmath}
\pdfoutput=1

\usepackage{cuted}
\usepackage{xcolor}

\begin{document}

\title{Observational test for $f(Q)$ gravity with weak gravitational lensing}

\author[0000-0003-0194-0697]{Qingqing Wang}
\affiliation{Department of Astronomy, School of Physical Sciences, University of Science and Technology of China, Hefei 230026, China}
\affiliation{CAS Key Laboratory for Researches in Galaxies and Cosmology, School of Astronomy and Space Science, University of Science and Technology of China, Hefei, Anhui 230026, China}
\affiliation{Deep Space Exploration Laboratory, Hefei 230088, China}

\author[0000-0002-5450-0209]{Xin Ren}
\affiliation{Department of Astronomy, School of Physical Sciences, University of Science and Technology of China, Hefei 230026, China}
\affiliation{CAS Key Laboratory for Researches in Galaxies and Cosmology, School of Astronomy and Space Science, University of Science and Technology of China, Hefei, Anhui 230026, China}
\affiliation{Deep Space Exploration Laboratory, Hefei 230088, China}
\affiliation{Department of Physics, Tokyo Institute of Technology, Tokyo 152-8551, Japan}

\author[0000-0003-0706-8465]{Yi-Fu Cai}
\affiliation{Department of Astronomy, School of Physical Sciences, University of Science and Technology of China, Hefei 230026, China}
\affiliation{CAS Key Laboratory for Researches in Galaxies and Cosmology, School of Astronomy and Space Science, University of Science and Technology of China, Hefei, Anhui 230026, China}
\affiliation{Deep Space Exploration Laboratory, Hefei 230088, China}

\author[0000-0003-1297-6142]{Wentao Luo}
\affiliation{Department of Astronomy, School of Physical Sciences, University of Science and Technology of China, Hefei 230026, China}
\affiliation{CAS Key Laboratory for Researches in Galaxies and Cosmology, School of Astronomy and Space Science, University of Science and Technology of China, Hefei, Anhui 230026, China}
\affiliation{Deep Space Exploration Laboratory, Hefei 230088, China}

\author[0000-0003-1500-0874]{Emmanuel N. Saridakis}
\affiliation{National Observatory of Athens, Lofos Nymfon, 11852 Athens, Greece}
\affiliation{Department of Astronomy, School of Physical Sciences, University of Science and Technology of China, Hefei 230026, China}
\affiliation{Departamento de Matem\'{a}ticas, Universidad Cat\'{o}lica del Norte, Avda. Angamos 0610, Casilla 1280 Antofagasta, Chile}

\footnote{Corresponding authors: yifucai@ustc.edu.cn; wtluo@ustc.edu.cn; msaridak@noa.gr}

\begin{abstract}
In this article we confront a class of $f(Q)$ gravity models with observational data of galaxy-galaxy lensing. Specifically, we consider the $f(Q)$ gravity models containing a small quadratic correction when compared with General Relativity (GR), and quantify this correction by a model parameter $\alpha$. 
To derive the observational constraints, we start by extracting the spherically symmetric solutions which correspond to the deviations from the Schwarzschild solution that depends on the model parameter in a two-fold way, i.e., a renormalized mass and a new term proportional to $r^{-2}$. Then, we calculate the effective lensing potential, the deflection angle, the shear component, and the effective Excess Surface Density (ESD)  profile. After that, we employ the group catalog and shape catalog from the SDSS DR7 for the lens and source samples respectively. Moreover, we handle the off-center radius as a free parameter and constrain it using the MCMC. Concerning the deviation parameter from GR we derive $\alpha=1.202^{+0.277}_{-0.179}\times 10^{-6} {\rm Mpc}^{-2}$ at 1 $\sigma$ confidence level, and then compare the fitting efficiency with the standard $\Lambda$CDM paradigm by applying the AIC and BIC information criteria. Our results indicate that the $f(Q)$ corrections  alongside off-center effects yield a scenario that is slightly favored.
\end{abstract}

\section{Introduction} \label{sec:intro}

The $\Lambda$-Cold Dark Matter ($\Lambda$CDM) paradigm, based on GR, on the Standard Model of Particle Physics, on cold dark matter and on the cosmological constant, has received support from a series of experimental observations and stands as the current Standard Model of 
Cosmology. However, with the advancement of observational techniques, the $\Lambda$CDM scenario seems to be   challenged by new observations, which may reveal possible tensions between its predictions and the data \citep{Perivolaropoulos:2021jda, Abdalla:2022yfr, Wong:2019,  
DiValentino:2020vvd, Yan:2019gbw}. These observational inconveniences, alongside the known theoretical problem of non-renormalizability of GR, led a significant amount of research to be devoted to the constructions of modified gravity \citep{CANTATA:2021ktz}.

The most direct way to obtain a modified gravity theory is to construct modifications starting from the Einstein-Hilbert Lagrangian, and extend it to $f(R)$ gravity \citep{Starobinsky:1980te, Capozziello:2002rd}, $f(G)$
gravity \citep{Nojiri:2005jg}, $f(P)$ gravity \citep{Erices:2019mkd}, and Lovelock gravity \citep{Lovelock:1971yv}, etc. Alternatively, one can 
construct gravitational modifications  by extending the underlying geometry itself. For instance, starting from the torsional formulation of gravity, namely the Teleparallel Equivalent of General Relativity (TEGR), one can extend 
it to $f(T)$ gravity \citep{Bengochea:2008gz, Cai:2015emx,Cai:2018rzd,Li:2018ixg,Ren:2021tfi,Ren:2022aeo,Hu:2023juh,Hu:2023xcf}, $f(T,{T}_{G})$ gravity \citep{Kofinas:2014owa, Kofinas:2014daa, Asimakis:2021yct}, $f(T,B)$ gravity 
\citep{Bahamonde:2015zma} etc.

One interesting class of gravitational modifications arise from the use of non-metricity. In particular, one can use a metric-incompatible affine connection with vanishing curvature and torsion, \citep{Nester:1998mp, BeltranJimenez:2017tkd} and construct the Symmetric Teleparallel Equivalent of General Relativity (STEGR), in which the Lagrangian is the non-metricity scalar $Q$. When extending the Lagrangian from $Q$ to a function $f(Q)$, one can obtain a novel class of theories \citep{Heisenberg:2023lru}. $f(Q)$ gravity possesses GR as a particular limit, however in general it can lead to different and interesting cosmological phenomenology
\citep{Lu:2019hra, Mandal:2020buf, Barros:2020bgg, DAmbrosio:2021zpm,  Dimakis:2021gby, Khyllep:2021pcu, Li:2021mdp, De:2022wmj,  Hohmann:2021ast, Kar:2021juu, Mandal:2021bpd, Quiros:2021eju, Wang:2021zaz, Albuquerque:2022eac,  Bahamonde:2022cmz, Bahamonde:2022zgj, Capozziello:2022tvv, Capozziello:2022zzh, DAgostino:2022tdk, Dimakis:2022wkj, Emtsova:2022uij, Lymperis:2022oyo, Narawade:2022cgb, Papagiannopoulos:2022ohv, Solanki:2022rwu,  Atayde:2023aoj, Bhar:2023zwi, Ferreira:2023awf, Jarv:2023sbp,  Koussour:2023rly, Mandal:2023cag, Maurya:2023szc, Mussatayeva:2023aoa, Najera:2023wcw, Paliathanasis:2023pqp,  Shabani:2023xfn, Sokoliuk:2023ccw,
Bhar:2024vxk, Capozziello:2024vix, Goncalves:2024sem,  Mhamdi:2024kgu, Calza:2022mwt, Hu:2022anq, Hu:2023gui}, while it can also be extended to $f(Q,C)$ gravity in order to incorporate boundary effects \citep{De:2023xua, Gadbail:2023mvu, Capozziello:2023vne, Junior:2024xmm}. Note that there is a discussion in the literature whether $f(Q)$ gravity suffers from strong coupling or the presence of ghosts \citep{Gomes:2023tur}, however one can find versions of the theory that are in general pathologies-free, for instance incorporating direct couplings of the matter ﬁeld to the connection or non-minimal couplings \citep{Heisenberg:2023lru, Heisenberg:2023wgk, DAmbrosio:2023asf}.

Although $f(Q)$ gravity has been confronted in detail with data at  
cosmological scales \citep{Anagnostopoulos:2021ydo, Lazkoz:2019sjl, Ayuso:2020dcu, Anagnostopoulos:2022gej, Yang:2024kdo, Yang:2024tkw, Capozziello:2022wgl}, up to now it has not been tested using data from intermediate scales. In particular, it is known that in the framework of GR, light is bent in the presence of a gravitational field, distorting the image of a celestial object located behind intervening matter, a phenomenon known as gravitational lensing \citep{Bacon:2000sy, Hoekstra:2008db, Luo:2017zbc, Cai:2023ite, Jiang:2024otl,Mo:2024rfq,Soares:2023uup,Soares:2023err}. This phenomenon can be utilized to investigate the characteristics of dark matter and dark energy on the galactic and cosmological scale, which applications include cosmic shear \citep{KiDS:2020suj}, galaxy-galaxy lensing \citep{BOSS:2016wmc} and CMB lensing \citep{Lewis:2006fu}. Therefore, by examining the distortion of images from a multitude of background galaxies, weak lensing can directly map the gravitational field of the foreground objects\citep{Salucci:2018hqu}, thus offering valuable insights into constraining modified gravity theories.

In \citep{Ren:2021} the authors calculated the lensing effects, such as deflection angles and magnifications, under covariant $f(T)$ gravity. Moreover, in \citep{Luo:2020} emergent gravity was investigated through galaxy-galaxy lensing where it was found that it is not capable to explain color dependence. Furthermore, \citep{Chen:2019} and \citep{Wang:2023qfm} utilized weak lensing signals to constrain $f(T)$ gravity by modeling the effective ESD in real and complex tetrads, respectively.

In the present work we are interested in performing a confrontation of $f(Q)$ gravity with galaxy-galaxy lensing data, in order to examine whether the theory can past the corresponding tests. The plan of the work is the following. In Section \ref{fttheory} we present $f(Q)$ gravity and we extract the spherically symmetric solutions. In Section \ref{wlensing} we briefly review the weak lensing mechanism and we apply it for the case of $f(Q)$ gravity, while we present the data that we use. Then, in Section \ref{Observationalconstraints} we perform the observational confrontation, 
extracting the corresponding constraints on the theory. Finally, in Section \ref{Conclusions} we conclude.

\section{Spherically symmetric solutions in $f(Q)$ gravity}\label{fttheory}

In this section we introduce $f(Q)$ gravity and we explore its application 
on spherically symmetric solutions.

\subsection{$f(Q)$ gravity}

Let us briefly review  the fundamentals and structure of gravity theories based on non-metricity. The general affine connection $\Gamma^{\alpha}_{\mu\nu}$ is expressible as
\begin{equation}
 \Gamma^{\alpha}_{\mu\nu}=\hat{\Gamma}^{\alpha}_{\mu\nu}+ K^{\alpha}_{\,\,\mu\nu}+L^{\alpha}_{\,\,\mu\nu} ~,
\end{equation}
where $\hat{\Gamma}^{\alpha}_{\mu\nu}$ denotes the Levi-Civita connection,
\begin{equation}
 K^{\alpha}_{\,\,\mu\nu} =\frac{1}{2}T^{\alpha}_{\,\,\mu\nu} +T^{\,\,\,\alpha}_{(\mu\,\,\,\nu)} 
\end{equation}
represents the contortion tensor with $T^{\alpha}_{\,\,\mu\nu}$  being the 
torsion tensor, and
\begin{equation}
 L^{\alpha}_{\,\,\mu\nu} = \frac{1}{2}Q^{\alpha}_{\,\,\mu\nu}-Q^{\,\,\,\alpha}_{(\mu\,\,\,\nu)}
\end{equation}
is the disformation tensor, which arises from the  non-metricity tensor defined as
\begin{equation}
    Q_{\alpha\mu\nu}\equiv\nabla_\alpha g_{\mu\nu} ~,
\end{equation}
where $g_{\mu\nu}$ is the metric tensor, and Greek indices cover the range of 
the coordinate space. 

Employing the general affine connection yields the 
torsion and curvature tensors as
\begin{eqnarray}
\label{Tortnsor}
 T^{\lambda}{}_{\mu\nu} &\equiv& 
 \Gamma^{\lambda}{}_{\mu\nu}\!-\!\Gamma^{\lambda}{}_{\nu\mu} ~,\\
 R^{\sigma}{}_{\rho\mu\nu} &\equiv& \partial_{\mu} \Gamma^{\sigma}{}_{\nu\rho}\! - 
 \!
 \partial_{\nu} \Gamma^{\sigma}{}_{\mu\rho}\! +\! \Gamma^{\alpha}{}_{\nu\rho} 
 \Gamma^{\sigma}{}_{\mu\alpha} \!- \!\Gamma^{\alpha}{}_{\mu\rho} 
 \Gamma^{\sigma}{}_{\nu\alpha} ~, \nonumber
\label{Rietsor}
\end{eqnarray}
and the non-metricity tensor as
\begin{eqnarray}
\label{NonMetrensor}
 Q_{\rho \mu \nu} \equiv \nabla_{\rho} g_{\mu\nu} = \partial_\rho g_{\mu\nu} - \Gamma^\beta{}_{\rho\mu} g_{\beta\nu} - \Gamma^\beta{}_{\rho\nu} g_{\mu\beta}  ~.
\end{eqnarray}

Setting non-metricity to zero recovers Riemann-Cartan geometry, with torsion nullified we obtain a torsion-free geometry, and with curvature nullified we acquire teleparallel geometry. On the other hand, making both non-metricity and torsion equal to zero (in this case the general connection is the Levi-Civita one) results in Riemann geometry, nullifying both non-metricity and curvature (using Weitzenb{\"{o}}ck connection) leads to Weitzenb{\"{o}}ck geometry, and setting both curvature and torsion to zero (employing symmetric teleparallel connection) yields symmetric teleparallel geometry \citep{BeltranJimenez:2017tkd, Jarv:2018bgs}.  

Gravity in Riemann geometry is described by a Lagrangian based on the Ricci 
scalar derived from curvature tensor contractions, leading to GR. Moreover, in Weitzenb{\"{o}}ck geometry, gravity is described by a Lagrangian based on the torsion scalar derived from torsion tensor contractions, forming the TEGR. Similarly, therefore, in symmetric teleparallel geometry, gravity is described via a Lagrangian formed by contractions of the non-metricity tensor, specifically the non-metricity scalar
\begin{equation}
\label{NontyScalar}
 Q = \frac{1}{2} Q_{\alpha\beta\gamma}Q^{\gamma\beta\alpha} -\frac{1}{4} Q_{\alpha\beta\gamma}Q^{\alpha\beta\gamma} +\frac{1}{4} Q_{\alpha}Q^{\alpha} -\frac{1}{2}Q_{\alpha}\tilde{Q}^{\alpha} \,,
\end{equation}
where
$Q_{\alpha}\equiv Q_{\alpha \ \mu}^{\ \: \mu} $ and
$\tilde{Q}^{\alpha} \equiv Q_{\mu }^{\ \: \mu \alpha} $.

Drawing inspiration from the modifications in $f(R)$ and $f(T)$ gravity, the 
Lagrangian for STEGR, namely $Q$, can be generalized to a function of choice, leading to the formulation of $f(Q)$ gravity. The action for this theory is given by \citep{BeltranJimenez:2017tkd}
\begin{equation}  
\label{fQaction}
 S =  -\frac{1}{2\kappa} \int {\mathrm{d}}^4 x \sqrt{-g}  f(Q) ~,
\end{equation}
where $g$ is the determinant of the metric and $\kappa$ is the gravitational 
constant. In this framework, the classical form of STEGR is recovered when the function $f(Q)$ is set to $Q$.

Variation of the  total action $S+S_m$, where $S_m$ is the action of the matter sector, gives rise to the field equations of $f(Q)$ gravity as
\citep{BeltranJimenez:2019tme, Dialektopoulos:2019mtr}: 
\begin{eqnarray}
&&  
\frac{2}{\sqrt{-g}} \nabla_{\alpha}\left\{\sqrt{-g} g_{\beta \nu} f_{Q} 
\left[- \frac{1}{2} L^{\alpha \mu \beta}+ \frac{1}{4} g^{\mu \beta} 
\left(Q^\alpha -  \tilde{Q}^\alpha \right) \right.\right.\nonumber\\
&&\left.\left. \ \ \ \ \ \ \ \ \ \ \ \ \ \ \ \ \ \  \ \ \ \ \ \ \ \ \ \ \ \, 
- \frac{1}{8} 
\left(g^{\alpha \mu} Q^\beta + g^{\alpha \beta} Q^\mu  
\right)\right]\right\}
 \nonumber \\
&& + f_{Q} \left[- \frac{1}{2} L^{\mu \alpha \beta}- \frac{1}{8} \left(g^{\mu 
\alpha} Q^\beta 
+ g^{\mu \beta} Q^\alpha  \right)
 \right. \nonumber\\
&&\left. \ \ \ \ \ \ \ + \frac{1}{4} g^{\alpha 
\beta} \left(Q^\mu -  \tilde{Q}^\mu \right)
\right] Q_{\nu \alpha 
\beta} +\frac{1}{2} \delta_{\nu}^{\mu} f=T_{\,\,\,\nu}^{\mu}\,,
\label{eoms}
\end{eqnarray}
where $f_{Q}=\partial f/\partial Q$ and $T_{\,\,\,\nu}^{\mu}$ is the matter 
energy momentum tensor.

\subsection{Spherically symmetric solutions}

Let us now proceed to the investigation of spherically symmetric solutions in 
the framework of $f(Q)$ gravity. Although the coincident gauge is widely used in the cosmological Friedmann-Robertson-Walker Cartesian coordinates, the spherical symmetry is not compatible with it. Therefore, we need to choose the affine connection corresponding to the spherical symmetric coordinate system. For the spherically symmetric we choose the diagonal form of the metric is:
\begin{equation}\label{eq:GeneralMetric}
ds^{2}=g_{tt}dt^{2}-g_{rr}dr^{2}-r^{2}d\Omega^{2} ~,
\end{equation}
here $g_{tt}$ and $g_{rr}$ are functions only of $r$.

Additionally, concerning the 
connection, one can perform the symmetry reduction imposing the additional assumptions that it is torsionless and stationary. We refer to \citep{Hohmann:2019fvf, DAmbrosio:2021zpm} for details of this reduction. After choosing the formula of $\Gamma^{r}{}_{\theta\theta}$, all the non-zero connection components can be driven by 
\begin{align}
\Gamma^{r}{}_{\theta\theta} &= \frac{r}{\sqrt{g_{rr}}} ~,\notag\\
\Gamma^{t}{}_{rr} &= -\frac{\Gamma^{t}{}_{\theta\theta}}{(\Gamma^{r}{}_{\theta\theta})^{2}}, ~
\Gamma^{r}{}_{rr} =\frac{-1-\partial_{r}\Gamma^{r}{}_{\theta\theta}}{\Gamma^{r}{}_{\theta\theta}},\notag\\
\Gamma^{r}{}_{\phi\phi} &=\textrm{sin}^{2}\theta\Gamma^{r}{}_{\theta\theta}, ~
\Gamma^{t}{}_{\phi\phi} =\textrm{sin}^{2}\theta\Gamma^{t}{}_{\theta\theta},\\
\Gamma^{\theta}{}_{\phi\phi} &=-\textrm{cos}\theta\textrm{sin}\theta,  ~\Gamma^{\phi}{}_{\theta\phi} =\textrm{cot}\theta, \notag\\
\Gamma^{\theta}{}_{r\theta} &=\Gamma^{\phi}{}_{r\phi} =-\frac{1}{\Gamma^{r}{}_{\theta\theta}},\notag
\end{align}
where $\Gamma^{t}{}_{\theta\theta}$ is an independent function.
Then the non-metricity scalar $Q$ is solvable and can be expressed as   
\citep{DAmbrosio:2021zpm} 
\begin{equation}\label{eq:QConstr}
 Q = \frac{2\left(1+\sqrt{g_{rr}}\right)\left(g_{tt}\,\left(1+\sqrt{g_{rr}}\right)+ r\,\partial_r g_{tt}\right)}{g_{tt}\, g_{rr}\, r^2} ~.
\end{equation} 

In order to proceed we must choose a specific $f(Q)$ form. In principle, we 
expect that every realistic modified theory of gravity should be a small 
deviation from GR. Hence, we deduce that every realistic $f(Q)$ theory ought to have the form  
\begin{equation}
 f(Q) = Q + \alpha\, Q^2 +{\cal{O}}\left(Q^3\right) ~,
\end{equation}
where $\alpha$ the model parameter. Thus, the spherically symmetric solutions can be approximated as
\begin{eqnarray}\label{Metricchoice}
 g_{tt} = g_{tt}^{(0)} + \alpha\, g_{tt}^{(1)} ~,~~
 g_{rr} = g_{rr}^{(0)} + \alpha\, g_{rr}^{(1)} ~,
\end{eqnarray}
where $g_{tt}^{(0)}$ and $g_{rr}^{(0)}$ correspond to the standard Schwarzschild solution, namely
\begin{eqnarray}\label{Schwarzschildsol}
 g_{tt}^{(0)} = -\Big( 1-\frac{2M}{r} \Big) ~,~~
 g_{rr}^{(0)} = -\frac{1}{g_{tt}^{(0)}} ~.
\end{eqnarray}
The general solution for the corrections can be extracted as 
\citep{DAmbrosio:2021zpm}   
\begin{eqnarray}
&&
g_{tt}^{(1)} = \Big( 1-\frac{2M}{r} \Big) c_2 + \frac{32}{3M^2} \Big(1-\frac{2M}{r} \Big)^\frac{3}{2} \nonumber\\
&&
+\frac{1}{M^2 r^3}
\Big[ \ln\big(1-\frac{2M}{r}\big) r^2 \big(r-3M\big) \nonumber\\
&& 
+M \Big( 2M^2 
+2r^2 +Mr \big( 12+c_1 r \big) \Big) \Big] \notag ~, \\
&&
g_{rr}^{(1)} = 
\frac{r}{(r-2M)^2} \Big[c_1-\frac{1}{M}\ln\left(1-\frac{2M}{r}
\right)+ \frac{46}{r}\notag \\
&&
-\frac{50 
M}{r^2} -\frac{16\sqrt{r-2M}(M-2r)(3M-r)}{3M\, 
r^\frac{5}{2}}\Big] ~,
\end{eqnarray}
with $c_1$, $c_2$ constants. For large $r$, the above solution can be further approximated as \citep{DAmbrosio:2021zpm}
\begin{align}
 g_{tt} &= -1+\frac{2M_\text{ren}}{r}+\alpha\frac{32}{r^2}\notag\\
 -\frac{1}{g_{rr}} &= -1+\frac{2M_\text{ren}}{r}+\alpha\frac{96}{r^2} ~,
\end{align}
with the renormalized mass
\begin{equation}
 2M_\text{ren} \equiv 2M -\alpha \Big( \frac{32}{3M}+c_1 \Big) ~.
\end{equation}
In summary, the approximate spherically symmetric solution for realistic $f(Q)$ theories is 
\begin{eqnarray}
\label{metric solution}
&& 
ds^{2}= 
-\left( 1-\frac{2M_\text{ren}}{r}-\alpha\frac{32}{r^2} \right) 
dt^{2} \nonumber \\
&&\!\!  \!\! \! 
+\left( 1-\frac{2M_\text{ren}}{r}-\alpha\frac{96}{r^2} \right)^{-1} dr^{2} + r^{2}d\Omega^{2} ~.
\end{eqnarray}

\section{Weak lensing theory and data}\label{wlensing}

In this section, we introduce the weak lensing mechanism. Weak lensing refers to the shearing of distant galaxy images due to the differential deflection of neighboring light rays. This effect results in a small signal, usually causing an ellipticity at a level of around $1\%$. Although this distortion is minor compared to the inherent shape of individual galaxies, it can be detected statistically by analyzing the coherence of the lensing shear across the sky \citep{Bartelmann:1999}. Therefore, the characteristics of spacetime, affected by the underlying theory of gravity, manifest in the lensing signal, providing a way to test the theory of gravity itself.

We are interested in exploring the effects on the light traveling from a distant celestial body, as it propagates through a foreground gravitational field on the way, within the framework of $f(Q)$ gravity. 
Based on the 
fact that photons undergo deflection upon encountering foreground gravitational fields, to identify and understand the differences between $f(Q)$ gravity and GR, a comparative study can be conducted by meticulously calculating the deflection angle, symbolized as $\hat{\alpha}$, within a static spherically symmetric spacetime.
The deflection angle can be calculated from the geodesic equation under a certain spherically symmetric spacetime.  The geodesic equation can be written using the corresponding affine connection and disformation tensor forms in non-metric gravity. According to the relation between affine connection and   Levi-Civitia connection, the geodesic equation can   be converted to the form under curvature gravity, which is essentially the same. Under modified gravity theories, the spherically symmetric metric solution is different, which results to the modification of the weak lensing formula.

The effective lensing potential is defined as \citep{Narayan:1996ba}:
\begin{equation}
 \Psi (\vec{\theta})=\frac{2 D_{ds}}{c^2D_s D_d}\int \Phi(D_d \theta,z) dz ~,
\end{equation}
where $D_d$, $D_{ds}$, and $D_s$ denote the angular diameter distances between the observer and the lens, the corresponding distance between the lens and source, and the one between the observer and the source. Moreover, the gravitational potential of the lens can be derived from the deflection angle as 
\begin{equation}
 \hat{\alpha} = \frac{2}{c^2}\int \nabla_{D_d \theta} \Phi(D_d \theta,z) dz ~.
\end{equation}
With the assumptions that the whole lensing system lies in the asymptotically 
flat spacetime regime, and that the distance of the light ray of closest 
approach $r_0$, 
as well as the impact parameter $b$, both lie outside the gravitational 
radius, the deflection angle for metric \eqref{metric solution} can be 
expressed as 
\citep{Keeton:2005}
\begin{align}
 \hat{\alpha}(r_{0})  
 &= 2\int^{\infty}_{r_{0}}\frac{1}{r^2}\sqrt{\frac{AB}{1/b^2-A/r^2}}dr -\pi ~,
\end{align}
where $A=g_{tt}$,$B=g_{rr}$, and the impact parameter defined by $b=\sqrt{\frac{r^2_0}{A(r_0)}}$ shows the perpendicular distance from the central axis of the lensing object to the 
asymptotic tangent of the light ray's trajectory when viewed with respect to an inertial observer.

Hence, for the spherically symmetric metric solution \eqref{metric solution}, 
the bending angle can be calculated under the parametrized-post-Newtonian 
(PPN) formalism, and can be expressed as a series expansion in the single 
quantity 
$\frac{m}{b}$ as \citep{Keeton:2005}
\begin{align}
\hat{\alpha}(b)&=A_{1}\left(\frac{m}{b}\right)+A_{2}\left(\frac{m}{b}
\right)^2+A_{3}\left(\frac{m}{b}\right)^3+\mathcal{O}\left(\frac{m}{b}\right)^4 \notag\\
 A_{1}&=4, 
 A_{2}=\frac{15}{4}\pi+\frac{40\alpha}{m^{2}}, 
 A_{3}=\frac{128}{3}+\frac{768\alpha}{m^{2}}  ~,
\end{align}
where we only wrote the coefficients of the first three orders. Therefore, we can acquire the three-dimensional gravitational potential as
\begin{equation}
 \Phi = -\frac{GM}{r} - \frac{15(GM)^{2}}{8r^{2}c^2}-\frac{20\alpha c^2}{r^{2}} 
~.
\end{equation}
Here, the first term represents the Newtonian potential, the second term 
constitutes the contribution from GR, and the third term 
embodies the modification arising from   $f(Q)$ gravity.

Subsequently, the shear tensor can be computed as a linear combination of the 
second partial derivatives of the potential, namely
\begin{align}
 \gamma_1 &\equiv \frac12\!\left(  \frac{\partial^2 \Psi}{\partial\theta_1^2} -\frac{\partial^2 \Psi}{\partial\theta_2^2} \right)  
\notag 
\\ &=
     -\frac{2D_{\mathrm{ds}}D_\mathrm{d}}{c^2 D_\mathrm{s}} \!
\left[\!\frac{2\Delta\Sigma(R)}{\pi}\!+\!\frac{30\pi 
\alpha c^2}{R^{3}}\!+\!\frac{45\pi (GM)^{2}}{16R^{3}c^2}\right]\!\cos2\phi
\end{align}
\begin{align}
    \gamma_2 &\equiv  \frac{\partial^2 \Psi}{\partial\theta_1\partial\theta_2} = \frac{\partial^2 \Psi}{\partial\theta_2\partial\theta_1} \notag \\  
    &=  -\frac{2D_{\mathrm{ds}}D_\mathrm{d}}{c^2 D_\mathrm{s}}\left[\!\frac{2\Delta\Sigma(R)}{\pi}\!+\!\frac{30\pi \alpha c^2}{R^{3}}\!+\!\frac{45\pi (GM)^{2}}{16R^{3}c^2}\right]\!\sin2\phi ~,
\end{align}
where $\Delta\Sigma(R)$ is the ESD under Newtonian approximation. Therefore, to account for differences, we define an effective ESD as follows,
\begin{align}
    \Delta \Sigma(R)_{\mathrm{eff}} &=\Sigma(\leq R)-\Sigma(R)=\gamma\Sigma_{\mathrm{crit}}  
\notag \\
    &=\Delta \Sigma(R)+\frac{15 \alpha c^2}{GR^{3}}+\frac{45 GM^{2}}{32R^{3}c^2} 
 ~,
\end{align}
where $\Sigma_{\mathrm{crit}}=\frac{c^{2}}{4\pi G}\frac{D_{\mathrm{s}}}{D_{\mathrm{ds}}D_\mathrm{d}}$ is the 
critical 
surface mass density. The quantity $\gamma = (\gamma_{1}^{2} + 
\gamma_{2}^{2})^{1/2}$ in the above expression describes the magnitude of the 
shear, since the shear components $\gamma_1$ and $\gamma_2$ introduce anisotropy (or astigmatism) shape distortion of the background galaxies, i.e. it quantifies how much the apparent shape of a galaxy has been altered compared to its intrinsic shape.

Having assembled all the requisite theoretical apparatus for the weak lensing 
phenomenon, let us now present the corresponding data. 
 We choose the data available within the Seventh Data Release of 
the SDSS DR7 \citep{SDSS:2008tqn}.
Positioned at Apache Point Observatory near Sacramento Peak in southern New 
Mexico, the SDSS leveraged a specialized 2.5-meter-wide-field telescope 
\citep{SDSS:2005} to capture CCD imagery across five broad bands photometry 
including the $u$, $g$, $r$, $i$, and $z$. 
The telescope employed twin instruments: a wide-field imager operating in drift 
scan mode to compile photometric catalogs,  and a dual multi-object 
spectrograph 
system feeding 640 fibers, delivering spectra spanning wavelengths from 3800 to 
9200 angstroms. These spectroscopic capabilities allowed for targeted analysis 
of celestial objects identified from the photometric data. 

The lenses sample for our study are selected from a galaxy group catalog built 
upon 
the spectroscopic data from the SDSS DR7 \citep{Yang:2007}, utilizing a 
sophisticated halo-based group detection algorithm originally developed in 
\citep{Yang:2006zf}. The robustness of this algorithm lies in its iterative 
process and the utilization of an adaptive filter tailored to the inherent 
characteristics of dark matter halos, while the halo mass in the catalog 
are 
approximated via abundance matching techniques. Following the estimation of the 
halo mass, further properties such as the velocity dispersion, virial radius, 
and other relevant metrics are inferred accordingly. This algorithm has 
undergone an enhancement \citep{Yang:2020eeb} to effectively handle galaxies 
with both photometric and spectroscopic redshifts, proving its efficacy in data 
analysis for the Dark Energy Spectroscopic Instrument (DESI) legacy imaging 
surveys DR8.
Based on this sample, we have selected lens galaxies suitable for testing 
gravitational theories. The various galaxies were identified by selecting 
information such as distance and redshift. Since we consider the gravitational 
potential generated by a spherically symmetric solution, the sample was selected 
from individual galaxy systems, leading to a 400,608   sample size.

We employ the shape catalog constructed by \citep{Luo:2016}, leveraging the SDSS 
DR7 imaging data as our source material, which comprises precise positional, 
morphological, 
shape uncertainty, and photometric redshift data for approximately 40 million 
galaxies, drawing upon the foundational work of \citep{Csabai:2007}. 
Furthermore,
the ESD $\Delta\Sigma(R)$ that we use   is measured through
  the weighted mean of source galaxy shapes as
\begin{align}
 \Delta\Sigma(R)= \frac{1}{2\Bar{R}}\frac{\Sigma 
(w_{i}e_{t}(R)\Sigma_{\mathrm{crit}})}{\Sigma w_{i}} ~,
\end{align}
where $w=\frac{1}{(\sigma_{\mathrm{shape}}^{2}+\sigma_{\mathrm{sky}}^{2})} $ is the corresponding 
weight,   with  $\sigma_{\mathrm{shape}}=<e^{2}>$   the shape noise and 
$\sigma_{\mathrm{sky}}=\frac{\sigma_{\mathrm{pix}}}{RF}\sqrt{4\pi n}$   the sky noise (with
$\sigma_{\mathrm{pix}}$   the galaxy size in pixels, $R$  the resolution factor, $F$ 
 the galaxy flux, and $n$  the sky and dark current in Analog-to-Digital Units.

\section{Observational constraints}
\label{Observationalconstraints}

\begin{figure*}
\centering
\includegraphics[width=15cm]{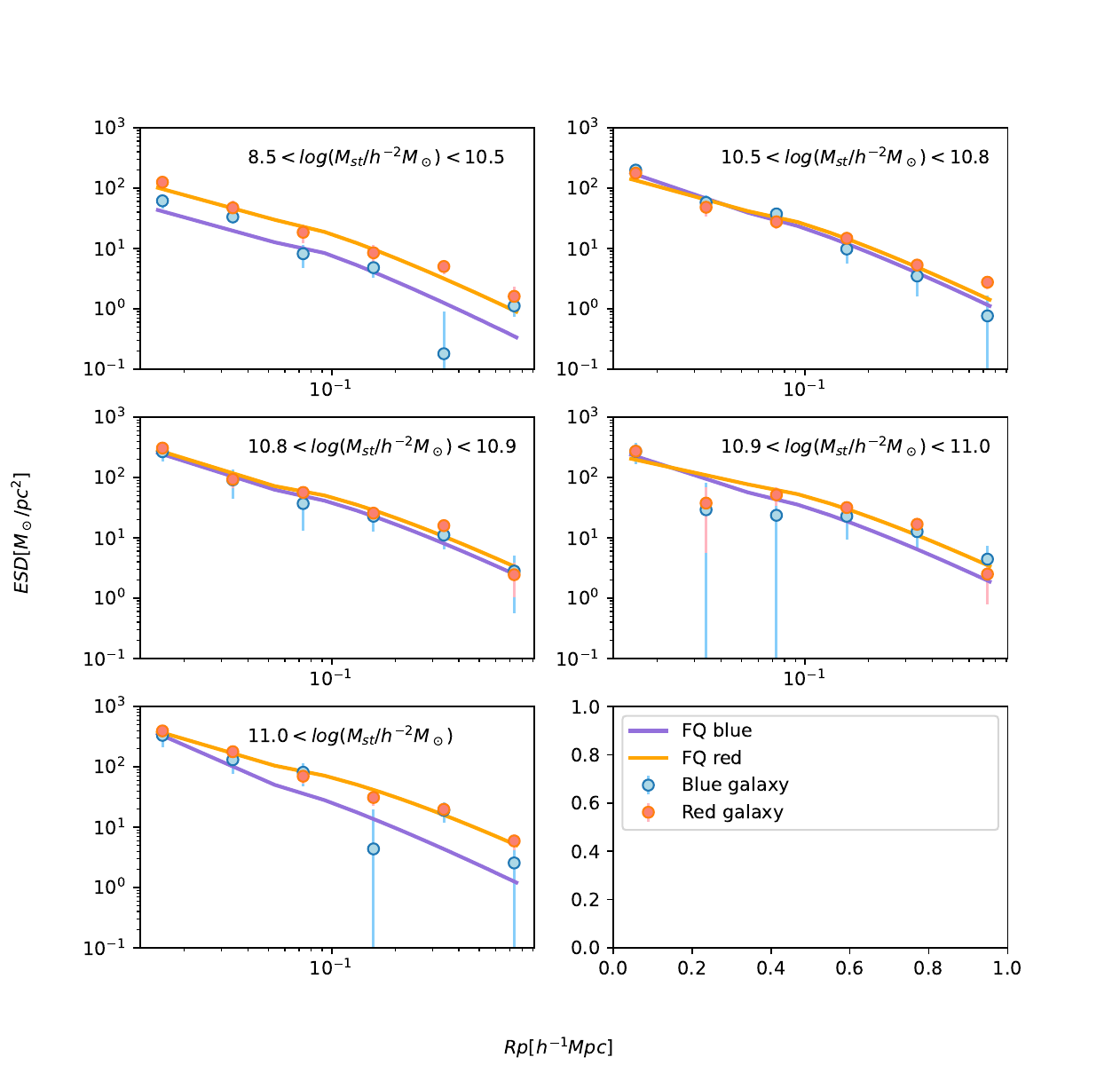}
\caption{{\it{The ESD profile and its best fits, as a function of the projected distance from the lens galaxy. The blue and orange points, 
alongside the corresponding errors, mark the weak lensing data for blue and red galaxies. For completeness, we have added the fit curves, shown with purple and orange lines. We have chosen $R_{\mathrm{sig}}=0.02$, which is the 
best fit value from MCMC constrain.}}}
\label{fig:ksquarefitting}
\end{figure*}

\begin{figure*}
\label{fig:Parameterconstrain}
\centering
\includegraphics[width=15cm]{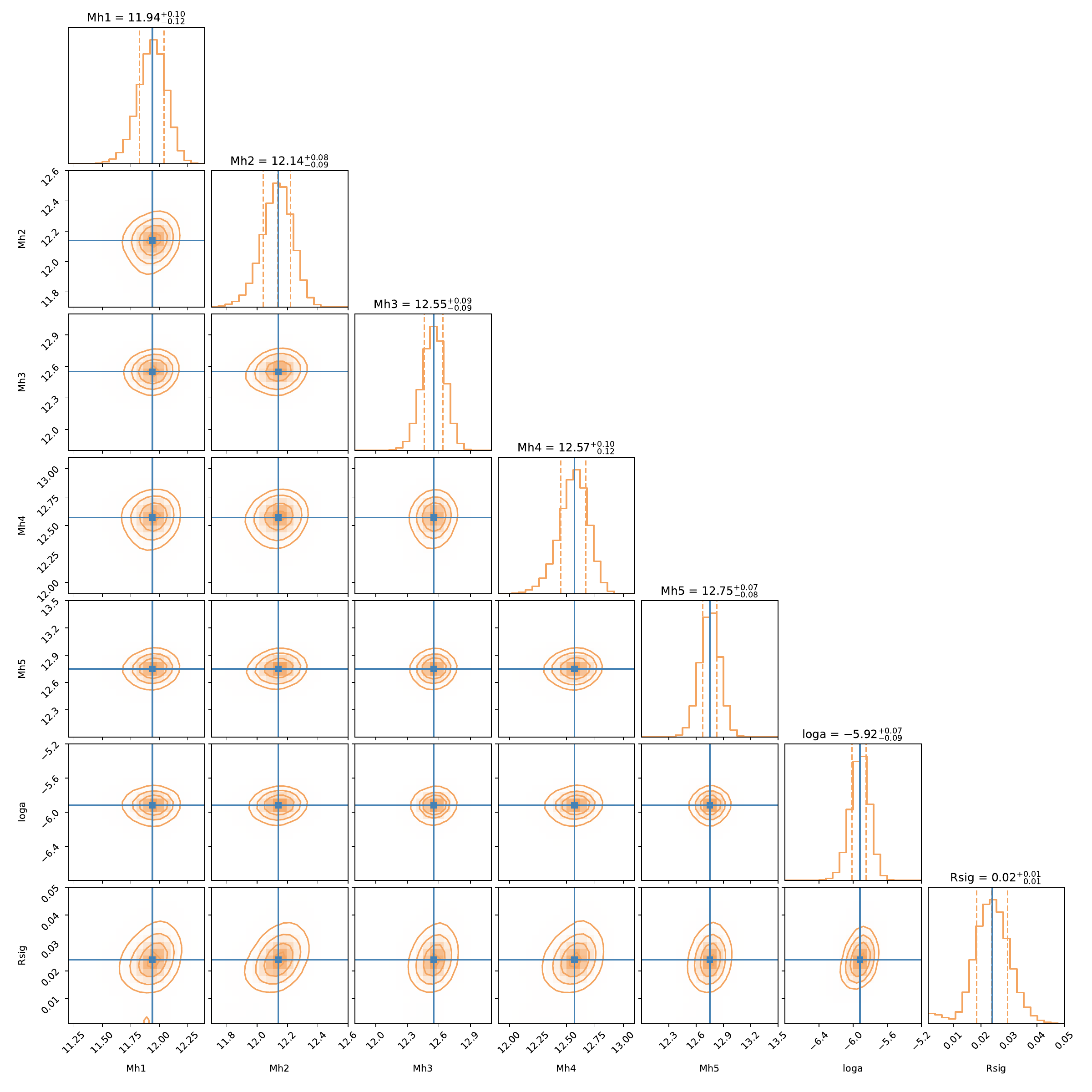}
\caption{{\it{The 1$\sigma$, 2$\sigma$ and 3$\sigma$, constraints for the halo mass deviation parameter and the off-center radius. 
}}}
\label{fig:constrain}
\end{figure*}

Let us now apply the steps described in the previous section, for the $f(Q)$ 
gravity. In particular, we desire to extract constraints on the model parameter $\alpha$ that quantifies the deviation from GR. Our approach 
consists of three main stages:
i) Model formulation: We adopt a modified NFW profile that incorporates both GR and additional $f(Q)$ corrections. This choice allows us to assess the impact of deviations from standard gravity on the inferred mass distribution. 
ii) Sample consideration: Recognizing that the modifications to the model parameters primarily manifest on smaller scales, we select our lens sample drawn from a single galaxy system, and in this case we can exclude the 
two-halo term from our analysis. This decision acknowledges the dominant role of host halo effects in the context of the chosen galaxy system and allows us to excavate the specific nature of the modified gravity under the galactic scale. 
iii) Off center reconstruction: The distribution of galaxies within the system serves as a proxy for the center of the gravitational potential. However, it is crucial to acknowledge the off-center effect, where the designated central galaxy may not precisely coincide with the true center of the gravitational potential. This misalignment introduces considerable uncertainty in the estimation of halo mass and the model parameter $\alpha$. In that case, we consider the off-center radius a free parameter and to be constrained during the MCMC process.

To evaluate ESD predictions our analysis employs ten catalogs of red and blue galaxies, divided into five mass intervals. We firstly employ the chi-square method to perform the quality-of-fit test of our model. Notably, the small stellar mass bin exhibits high signal-to-noise ratios and the $f(Q)$ correction significantly impacts the low mass regime, contributing to reduced $\chi^{2}$. Fig.~\ref{fig:ksquarefitting} reveals that the poor signal quality of blue galaxies in the large mass bin leads to decreased information content in their ESD profiles, while the red galaxies generally display superior signal quality. Therefore, we combine the data from red and blue galaxies, and we employ the concentration-mass relation from \citep{Neto:2007vq} as a Gaussian prior to mitigate degeneracies with other model parameters. Then we use the MCMC method to derive the final parameter space for both halo properties and model parameter $\alpha$.

Performing the above steps, we eventually extract the constraint for the $\alpha$ parameter, which is
\begin{align}
 \alpha=1.202^{+0.277}_{-0.179}\times 10^{-6} \rm Mpc^{-2} ~,
\end{align}
at 1$\sigma$ confidence level. Moreover, for the off-center radius we obtain $R_{\mathrm{sig}}=0.02\rm Mpc$. Finally, in Fig.~\ref{fig:constrain} we present the two-dimensional fitting posterior, and the constraints for various quantities.
Interestingly enough, we deduce that the parameter $\alpha$ is constrained to 
small, but non-zero values, which implies that $f(Q)$ corrections on top of 
GR are favoured by the weak lensing data.

In order  to assess the efficiency of the fittings in comparison with 
$\Lambda$CDM paradigm, we employ the Akaike Information 
Criterion (AIC) and the Bayesian Information Criterion (BIC). These statistical measures allow us to quantify the trade-off between model complexity and its ability to explain the observed data.
The AIC \citep{Akaike:1974vps} is derived from the Kullback-Leibler 
information, serving as an asymptotically unbiased estimator of model fitness. 
It is given by:
\begin{align}
 AIC \equiv -2ln\mathcal{L}_{\mathrm{max}}+2p_{\mathrm{tot}} ~, 
\end{align}
where $\mathcal{L}_{\mathrm{max}}$ denotes the maximum likelihood achieved by the model 
when fitted to the data, and $p_{\mathrm{tot}}$ represents the total number of free 
parameters in the model.
The BIC \citep{Kass:1995loi} is related to the Bayesian evidence and is 
defined as:
\begin{align}
 BIC \equiv -2ln\mathcal{L}_{\mathrm{max}}+p_{\mathrm{tot}}ln(N_{\mathrm{tot}}) ~,
\end{align}
where $N_{\mathrm{tot}}$ is the number of samples. Finally,  based on their information 
criterion 
differences $\Delta\text{IC}$, and using 
Jeffreys' classification \citep{Anagnostopoulos:2019miu}, we compare  $f(Q)$ 
gravity and  $\Lambda$CDM scenario.

\begin{table} 
\begin{center}
\begin{tabular}{c|c|c|c|c}
\hline\hline
Model  & $BIC$ & $AIC$ &$\Delta BIC$ & $\Delta AIC$ \\  
\hline
$\Lambda$CDM & 61.417 & 54.411& 0 & 0 \\   
$\Lambda CDM + off$& 65.507 & 57.100 & 4.090 & 2.689 \\
$f(Q)$ & 64.681 & 56.274 & 3.264 & 1.863\\
$f(Q) + off$ & 50.447 & 40.639 & -10.970 & -13.772 \\

\hline\hline
\end{tabular}
\end{center}
\caption{The information criteria BIC and AIC for the standard 
NFW model in 
$\Lambda$CDM scenario, and the  effective one in the framework of $f(Q)$ 
gravity, alongside their corresponding 
differences. We have performed our analysis  
without and with the off-center effect.}
\label{tab:bic}
\end{table}

We consider  two cases. Firstly, without considering 
 off-center effect, and secondly considering the off-center 
distance dispersion given by the best value $R_{\mathrm{sig}}=0.02\rm Mpc$. The obtained 
results are summarized in Table \ref{tab:bic}.
As we can see, in the absence of off-center effects, the fitting procedure
 indicates a reduced fit quality or increased complexity, which makes  
$\Lambda$CDM scenario more favourable. Nevertheless, in the presence of off-center effects, the information criteria analysis reveals a substantial improvement for $f(Q)$ gravity, thus it is favored in this case compared with the $\Lambda$CDM paradigm.
Hence, 
the combination of $f(Q)$ gravity  alongside off-center effects yields 
the best balance between model fit and complexity, making it the most favored 
model according to both information criteria. 

\section{Conclusions}
\label{Conclusions}

In this work we performed  a confrontation of $f(Q)$ gravity with observations 
at   galactic scales, namely using  galaxy-galaxy lensing data. We considered a 
small quadratic $f(Q)$ correction on GR, quantified by the new
model parameter $\alpha$, an ansatz  which is expected to hold in every 
realistic modification of gravity, and then we extracted the resulting 
spherically symmetric solutions. As we saw, these correspond to a deviation from 
Schwarzschild solution that depends on  $\alpha$ in a two-fold way, namely a     
renormalized mass and a new term proportional to $r^{-2}$.

As a next step we calculated the effective lensing potential as well as the 
deflection angle, and thus we extracted the correction  on the shear component, 
and effective ESD profile. As we saw, the  $r^{-2}$ correction  amplifies the 
effect of weak lensing, predominantly manifesting at small scales.
We employed the group catalog and shape catalog from the SDSS DR7 for the 
lens and source samples respectively. Lens galaxy samples were segregated into 
blue star-forming galaxies and red passive galaxies and then divided into five 
stellar mass bins to investigate the model dependence on galaxy color. Modeling 
the ESD with the NFW profile containing $f(Q)$ correction terms, we found a 
minimal dependence of the modified gravitational model on galaxy color. 

Furthermore, in order to investigate the influence of the off center effect on 
the parameter 
estimation, we set the off center radius to be an free parameter and try to 
constrain it  together with other model and halo properties using the 
MCMC program. Finally, we extracted the most recent estimation for the 
deviation 
parameter as $\alpha=1.202^{+0.277}_{-0.179}\times 10^{-6} \rm Mpc^{-2}$ at   1 
$\sigma$ confidence level, with the off-center radius estimation as 
$R_{\mathrm{sig}}=0.02^{+0.01}_{-0.01}\rm Mpc$. 
To our knowledge, this is the first constraint for $f(Q)$ gravity at galactic 
scales. The fact that the  zero value is excluded at 1 $\sigma$ confidence 
interval, indicates that small corrections on top of GR are 
favored.

In order to rigorously evaluate the fitting efficiency of $f(Q)$ gravity 
comparing to that of $\Lambda$CDM scenario,    we calculated the AIC and BIC 
information criteria values with and without considering the off-center effect. 
Considering the importance of the intrinsic off-center effect of the galaxy 
sample to constrain   $f(Q)$ gravity, we chose the best fit value of the 
off-center radius for the calculation. Our analysis revealed that $f(Q)$ gravity 
demonstrates a heightened congruence with the galaxy-galaxy lensing data when 
the off-center effect is appropriately taken into account. Interestingly 
enough, $f(Q)$ corrections  alongside off-center effects yield a scenario that 
is slightly favoured over  $\Lambda$CDM paradigm, and hence can challenge the 
latter.

Since the contribution of modified gravity is more pronounced at small scales, 
the effects of baryonic feedback effects \citep{Semboloni:2011fe, Duffy:2010hf} 
at that scale may show up in different samples of galaxies. In that case,  we 
acknowledged that a comprehensive joint analysis incorporating all relevant 
astrophysical influences would likely  enhance the precision of our findings, 
constituting a valuable pursuit for future investigations. Additionally, our 
analysis tested $f(Q)$ gravity at   galaxy scales, yet validation at smaller 
solar-system scales and larger cluster/filament scales is imperative, 
especially since future research is expected to leverage strong or 
filamentary gravitational lensing. Finally, we mention that our methodology is 
applicable to other extended gravity theories, at low-redshifts and weak-field 
contexts, and thus exhibits considerable potential for broad implementation in 
the wealth of data from upcoming galaxy surveys.

\section{Acknowledgments}
We thank Geyu Mo, Di He, Yuhang Yang, Bo Wang, Qinxun Li, Yumin Hu, Hongsheng Zhao for valuable discussions. This work is supported in part by National Key R\&D Program of China (2021YFC2203100), by NSFC (12433002, 12261131497), by CAS young interdisciplinary innovation team (JCTD-2022-20), by 111 Project (B23042), by Fundamental Research Funds for Central Universities, by CSC Innovation Talent Funds, by USTC Fellowship for International Cooperation, by USTC Research Funds of the Double First-Class Initiative. ENS acknowledges the contribution of the LISA CosWG and the COST Actions CA18108 “Quantum Gravity Phenomenology in the multi-messenger approach'' and CA21136 “Addressing observational tensions in cosmology with systematics and fundamental physics (CosmoVerse)''. 
We acknowledge the use of computing clusters {\it LINDA}  \& {\it JUDY} of the particle cosmology group at USTC.

\software{astropy \citep{2013A&A...558A..33A,2018AJ....156..123A}}

\bibliography{ftlensing}{}
\bibliographystyle{aasjournal}



\end{document}